\documentclass[aps]{revtex4}
 
\usepackage{amsmath}
\usepackage{amssymb}

\newcommand{\beq}{\begin{equation}}
\newcommand{\eeq}{\end{equation}}
\newcommand{\eps}{\epsilon}
\newcommand{\curl}{\nabla\times}
\newcommand{\vv}{\mathbf{v}}
\newcommand{\ups}{\Upsilon}
\newcommand{\lap}{\Delta}
\newcommand{\ilap}{\Delta^{-1}}

\newcommand{\ben}{\begin{eqnarray}}
\newcommand{\een}{\end{eqnarray}}
\newcommand{\benn}{\begin{eqnarray*}}
\newcommand{\eenn}{\end{eqnarray*}}
\newcommand{\tn}{\tilde{n}}

\def\pd#1#2{\frac{\partial #1}{\partial #2}}

\begin{document}

\title{Hamiltonian derivation of the Charney-Hasegawa-Mima equation}

\author{E. Tassi$^1$, C. Chandre$^1$, P.J. Morrison$^2$}
\affiliation{$^1$ Centre de Physique Th\'eorique, CNRS -- Aix-Marseille Universit\'es, Campus de Luminy, case 907, F-13288 Marseille cedex 09, France \\ $^2$ Institute for Fusion Studies and Department of Physics, The University of Texas at Austin, Austin, TX 78712-1060, USA}
\date{\today}
\baselineskip 24 pt

\begin{abstract}
The Charney-Hasegawa-Mima equation is an infinite-dimensional Hamiltonian system with dynamics generated by a noncanonical Poisson bracket.  Here a  first principle Hamiltonian derivation of this system, beginning  with  the ion fluid dynamics and its known Hamiltonian form, is given.
\end{abstract}
\maketitle

\section{Introduction}

When dissipative terms are dropped, all of the important models of  plasma physics are described by  partial differential equations that possess Hamiltonian form in terms of noncanonical Poisson brackets.  For example, this is the case for  ideal magnetohydrodynamics~\cite{morr80a,Mor84,mars84},  the Vlasov-Maxwell equations~\cite{morr80b,mars82,bial84}, and other systems (see Refs.~\cite{morr82,morr98,morr05} for review).   Among these,  there exist several reduced fluid models whose Hamiltonian structure has been derived {\it a posteriori}. These include the four-field model for tokamak dynamics of Hazeltine et al.~\cite{Haz87};    models for collisionless magnetic reconnection derived and investigated by Schep et al.~\cite{Sch94},  Kuvshinov et al.~\cite{Kuv94},  and Tassi et al.~\cite{tassi08};  and  the recent gyrofluid model  of  Waelbroeck et al.~\cite{Wae09}.  The noncanonical Hamiltonian formulation has also been adopted to investigate the electron temperature gradient driven mode \cite{Gue04} and convective-cell formation in plasma  fluid systems \cite{Kro04}.  In addition to these fluid models, the Hamiltonian structure of kinetic and reduced kinetic equations has also been highlighted, for example,  in   guiding-center theory and gyrokinetics (see Refs.~\cite{Lit79,Lit81,Lit82,briz07,briz09} for review).   

 This Hamiltonian form  originates from the Hamiltonian and action principle forms of the basic electromagnetic interaction, i.e.,  the Hamiltonian form possessed by the equations that describe a system of charged particles coupled to Maxwell's equations (see, e.g., Ref.~\cite{morr05} for discussion).   It is now well established that there exist numerous advantages of such a Hamiltonian formulation, among which are the identification of conserved quantities (that are important for the verification of numerical codes), the study of stability, the use of techniques for Hamiltonian systems  like averaging and perturbation theory, etc.  Here we perform a perturbative derivation  within the noncanonical Hamiltonian context, which means the Poisson bracket as well as the Hamiltonian must be expanded.

In a nutshell, a Hamiltonian system is a system whose dynamics of any observable $F$ (depending on a finite or infinite number of variables) can be written using a Hamiltonian (scalar) function $H$ and a Poisson bracket $\{\cdot ,\cdot\}$ as
$$
\frac{\partial F}{\partial t}=\{F, H\},
$$
where the Poisson bracket satisfies the following properties: bilinearity, antisymmetry, Leibniz rule, and Jacobi identity.
Given a reduced model whose dynamics is given by a partial differential equation, it is in general  difficult to guess whether or not the model
 is a Hamiltonian system, and if it is, finding the Hamiltonian and the Poisson bracket may be similarly difficult. There are basically two methods for finding Hamiltonian structure:   the first method is to use physical intuition to obtain the Hamiltonian (energy) and   to construct a general class of antisymmetric operators which, when acting  on the gradient of the Hamiltonian, produces the equations of motion.  Then, the Jacobi identity is used to select from the class the desired operator that is the essence of the noncanonical   Poisson bracket.    This method has been used to obtain a large number of basic and approximate Poisson brackets for fluid and plasma dynamics, examples being the reduced fluid models  cited above.  The second method begins from a known or postulated  action principle, in the latter case obtained by   using physical intuition to  obtain the `energies' of the Lagrangian.  Usually associated with the action principle  is a  canonical Hamiltonian 
 description,  which can be written by means of the chain rule in terms of physical variables of interest (e.g. Refs.~\cite{mars82,morr05}) resulting in a noncanonical Poisson bracket.

If one begins from some Hamiltonian parent model, some   basic starting point in the derivation,  and introduces  crude approximations suggested, e.g., by  physical considerations of some experimental set-up, then the Hamiltonian structure can be easily  destroyed. The Hamiltonian form of the resulting system must therefore be verified, in particular,  the Jacobi identity for the Poisson bracket.  Given this verification, the reduced model is naturally equipped with a Hamiltonian structure since the Poisson bracket and the Hamiltonian function are provided by the derivation process (for an example of this derivation process, see Ref.~\cite{bach08}).

In this paper we consider the derivation of the Charney-Hasegawa-Mima (CHM) equation~\cite{char71,hase77}, which describes both the dynamics of Rossby waves of geophysical fluid dynamics  (see, e.g., Ref.~\cite{ped79}) and drift waves in inhomogeneous plasmas  (see, e.g.,  Ref.~\cite{Hor99}).  We focus on the derivation in the plasma physics context but our analysis can be easily adapted to the geophysical context.  In particular, we show how the Hamiltonian structure is preserved in the derivation of the CHM equation starting from a fluid parent model. In the present approach the Hamiltonian structure is provided by the derivation process, and the Jacobi identity need not be checked.    

We consider a plasma under the influence of a constant and uniform magnetic field ${\bf B}=B\hat{\bf z}$. The relevant dynamics occurs in the (two-dimensional) transverse plane whose coordinates in a given basis are denoted by $x$ and $y$. Under some assumptions, the CHM  equation gives the following evolution of the electrostatic potential $\phi(x,y,t)$ generated by the plasma:
\beq
\frac{\partial}{\partial t} (\phi-\lap \phi)=\left[\phi, \lap \phi +\lambda \right],
\eeq
where the bracket $[\cdot,\cdot]$ is given by
$$
[f,g]=\pd{f}{x}\pd{g}{y}-\pd{f}{y}\pd{g}{x}=\hat{\bf z}\cdot \nabla f\times \nabla g,
$$
and $\lambda$ is any function of $x$ and $y$ (related to the equilibrium configuration). The infinite-dimensional phase space is composed of  the variables $\phi(x,y)$ for any point $(x,y)$ in the transverse plane. The space of observables, $F$, for this system is composed of  functionals of $\phi$.
It has been shown in Ref.~\cite{wein83} that this equation possesses an infinite-dimensional Hamiltonian structure
where the Hamiltonian is
$$
H(\phi)=\frac{1}{2}\int d^2 x \left(\phi^2+\vert \nabla \phi\vert^2 \right),
$$
and the noncanonical Poisson bracket is 
\ben
\{F,G\}=-\int d^2x (\phi-\lap\phi-\lambda)\left[(1-\lap)^{-1}F_\phi,(1-\lap)^{-1}G_\phi \right],
\een
where $F_\phi$ denotes the functional derivative of the functional $F$ with respect to the variable $\phi$.
This Hamiltonian structure was found  {\em ad hoc} in Ref.~\cite{wein83} by  an educated guess in  analogy with the vorticity equation for two-dimensional incompressible flow (see e.g. Ref.~\cite{morr82}). This analogy is rather straightforward if we consider the dynamics for the field $q=\lap\phi - \phi+\lambda$ which is given by the Hamiltonian
$$
H=\frac{1}{2}\int d^2x (q-\lambda)(1-\lap)^{-1}(q-\lambda),
$$
and the Lie-Poisson bracket
$$
\{F,G\}=\int d^2x \, q[F_q,G_q],
$$
which is of the same form as that for the Vlasov-Poisson system \cite{morr80b} and a quite general class of systems \cite{morr03}.  In what follows, we start by considering a Hamiltonian formulation for the fluid equations for the ions (in Sec.~\ref{sec:ion}) and derive the above Hamiltonian and Poisson bracket from this formulation (in Sec.~\ref{sec:HM}).

\section{Ion fluid equations as a Hamiltonian system}
\label{sec:ion}

We start the derivation of the CHM  equation from two dynamical equations: one describing the transverse dynamics of the ion velocity field $\vv(x,y,t)$ and the other describing the dynamics of the ion density field $n(x,y,t)$:
\ben
&& M\left(\dot{\vv} +(\vv \cdot \nabla) \vv\right)=-e\nabla\phi+e\vv \times {\bf B},\label{eqn:s1}\\
&& \dot{n}=-\nabla\cdot (n\vv),\label{eqn:s2}
\een
where the dot indicates the partial derivative with respect to time $t$.
The electrostatic  potential $\phi$ is obtained from the dynamics of the electrons: by neglecting their inertia,   the electron density obeys the Boltzmann law
\beq
n_e=n_0 \exp\left(e\phi/T\right),
\eeq
where $T$ is the electron temperature and $n_0=n_0(x,y)$ is the electron density at equilibrium. From the quasi-neutrality condition, we obtain that $n=n_e$. The total energy of the ions,  given by the sum of their kinetic energy plus the potential energy provided by the electric field,  is a conserved quantity that is also a good candidate for the Hamiltonian of the system  of  Eqs.~(\ref{eqn:s1}-\ref{eqn:s2}). This Hamiltonian is written as
\beq \label{ham}
H(n,\mathbf{v})=\int d^2 x \left[n\frac{v^2}{2}+\frac{T}{M}n\left(\ln\left(\frac{n}{n_0}\right)-1\right)\right].
\eeq
The dynamics is determined by the Poisson bracket
\beq \label{br1}
\{F,G\}=-\int d^2 x \left[F_{\mathbf{v}}\cdot \nabla G_n - G_{\mathbf{v}}\cdot\nabla F_n-\left(\frac{\curl \bf{v}}{n}+\frac{\omega_c}{n}\hat{\bf z}\right)\cdot(F_{\mathbf{v}}\times G_{\mathbf{v}})\right],
\eeq
where $\omega_c=e B/M$.  The bracket of (\ref{br1}) is identical to a portion of that of Ref.~\cite{morr80a}  with the inclusion of an additional `vorticity' term, $\omega_c\hat{\bf z}/n$; consequently, it is known to satisfy the Jacobi identity.  It is easy to verify that the ion momentum equation is obtained from the bracket of the  velocity field with the Hamiltonian~(\ref{ham}):
$$
\dot{\vv}\equiv \{\vv, H\}=-(\vv \cdot\nabla)\vv -\frac{T}{M}\nabla \ln\left(\frac{n}{n_0}\right)+\omega_c \vv \times \hat{\bf z},
$$
and, similarly,  the ion continuity equation is given by
$$
\dot{n}\equiv \{n,H\}=-\nabla\cdot(n\vv).
$$

\section{Charney-Hasegawa-Mima equation}
\label{sec:HM}
 
Without loss of generality we write the vector field $\vv(x,y,t)$ in terms of two scalar fields $\phi$ and $\ups$ as
\beq \label{vel}
\vv=\hat{\bf z} \times\nabla\phi+\nabla\ups,
\eeq
where one function is related to $\nabla\cdot\vv$ and the other to $\nabla\times\vv$ by the relations: $\Delta \phi=\hat{\bf z}\cdot \nabla \times \vv$ and $\Delta \ups=\nabla\cdot\vv$. In fact, we find it more convenient to consider a related  change of variables~$(n,\vv)\mapsto (\tn,q,D)$ defined by
\benn
&& \tn=n,\\
&& q=\frac{\hat{\bf z}\cdot\nabla\times \vv+\omega_c}{n},\\
&& D=\nabla\cdot \vv.
\eenn
The above equations are incomplete because they do not possess a unique  inverse.   However, a unique inverse is  defined by the following:
\benn
&& n=\tn,\\
&& \vv =\hat{\bf z}\times \nabla \ilap (q\tn-\omega_c)+\nabla\ilap D,
\eenn
where
$$
\ilap F=-\frac{1}{2\pi}\int d^2x' \ln \Vert {\bf x}-{\bf x}' \Vert F({\bf x}').
$$
In terms of the new variables $(\tn,q,D)$, the Hamiltonian~(\ref{ham}) becomes
\ben
H(\tn,q,D)=\int d^2 x && \left[\tn\left(\frac{|\nabla\ilap (q\tilde{n}-\omega_c)|^2}{2}+[\ilap (q\tilde{n}-\omega_c),\ilap D]\right.\right. \nonumber\\
&& \left.\left.\quad +\frac{|\nabla\ilap D|^2}{2}\right)+\frac{T}{M}\tn\left(\ln\left(\frac{\tn}{n_0}\right)-1\right)\right],
\een
since $\vert \hat{\bf z} \times \nabla f\vert^2=\vert \nabla f\vert^2$ for any function $f$ of $x$ and $y$, and  the bracket~(\ref{br1}) becomes
\ben \nonumber
\{F,G\}=-\int d^2 x && \left(-\nabla F_D\cdot\nabla G_{\tn} + \nabla G_D\cdot \nabla F_{\tn}+\frac{G_q}{\tn}\nabla F_D\cdot\nabla q-\frac{F_q}{\tn}\nabla G_D\cdot\nabla q\right. \nonumber \\
&& \left.\quad -q\left[\frac{F_q}{\tn},\frac{G_q}{\tn}\right]-q[F_D,G_D]\right). \nonumber
\een

It should be noted that Casimir invariants of such a bracket, which are the functionals that Poisson commute with all the other functionals ($\{C,G\}=0$ for all functionals $G$), are given by
$$
C=\int d^2x\,  \tn \mathcal{F}(q),
$$
where $\mathcal{F}$ is any function of $q$.

We first assume that the variables evolve slowly with time, which is equivalent to adding a factor of  $1/\eps$  in front of the Hamiltonian, 
\benn
H(\tn,q,D)=\frac{1}{\eps}\int d^2 x && \left[\tn\left(\frac{|\nabla\ilap (q\tilde{n}-\omega_c)|^2}{2}+[\ilap (q\tilde{n}-\omega_c),\ilap D]\right.\right.\\
&& \left.\left.\quad +\frac{|\nabla\ilap D|^2}{2}\right)+\frac{T}{M}\tn\left(\ln\left(\frac{\tn}{n_0}\right)-1\right)\right] , 
\eenn
then we introduce an  $\eps$-ordering  for the dynamical variables. The hypothesis is that the system of interest is near an equilibrium state whose spatial variations are of order $\eps$:
\benn
&& n({\bf x},t)=n_0(\eps{\bf x}) +\eps n_1({\bf x},t),\\
&& \vv({\bf x},t)=\eps \vv_1 ({\bf x},t),
\eenn
which translates into an assumption on the new variables $(\tn,q,D)$ and, in particular,  on  the definition of new dynamical variables $(\tn_1,q_1,D_1)$, 
\benn
&& \tn=\tn_0+\eps \tn_1,\\
&& q=q_0+\eps q_1,\\
&& D=\eps D_1,
\eenn
where $q_0=\omega_c/\tn_0$ and $\tn_0=n_0(0,0)$ are constant (the spatial variations of $n_0$ are included in $\tn_1$).  Notice that the potential energy can be rewritten as
$$
\tn\left(\ln\left(\frac{\tn}{n_0}\right)-1\right)=\tn\left(\ln\left(\frac{\tn}{\tn_0}\right)-1\right)-\tn\ln\frac{n_0}{\tn_0},
$$
with the following expansion:
$$
\tn\left(\ln\left(\frac{\tn}{n_0}\right)-1\right)=-\tn_0-\tn_0\ln\frac{n_0}{\tn_0}+\eps^2\frac{\tn_1^2}{2\tn_0}-\eps\tn_1 \ln\frac{n_0}{\tn_0}+O(\eps^3).
$$
The term $-\eps\tn_1\ln(n_0/\tn_0)$ is of order $\eps^2$,  due to the spatial variations of $n_0$,  which can be seen by writing $n_0=\tn_0+\eps\delta n_0$:
$$
-\eps\tn_1 \ln\frac{n_0}{\tn_0}=-\eps^2 \frac{\tn_1 \delta n_0}{\tn_0}+O(\eps^3).
$$
Next, we expand the Hamiltonian and the Poisson bracket: the Hamiltonian is
$$
H=\eps \int d^2x\,  \tn_0\left[ \frac{\vert\nabla\ilap(q_1\tn_0+q_0\tn_1)\vert^2}{2}+\frac{\vert\nabla\ilap D_1\vert^2 }{2}+\frac{T}{2M}\frac{\tn_1^2-2\tn_1 \delta n_0}{\tn_0^2}\right]+O(\eps^2),
$$
since $\int d^2x \, \tn_0 [\ilap(q_1\tn_0+q_0\tn_1),\ilap D_1]=0$,
and the Poisson bracket is  
\benn
\{F,G\}&=& \frac{1}{\eps^2}\int d^2x \left(\nabla F_{D_1}\cdot \nabla G_{\tn_1}-\nabla F_{\tn_1}\cdot \nabla G_{D_1} \right)\\
&& -\frac{1}{\eps} \int d^2x \left( \frac{G_{q_1}}{\tn_0}\nabla F_{D_1}\cdot\nabla q_1-\frac{F_{q_1}}{\tn_0}\nabla G_{D_1}\cdot\nabla q_1 -q_1\left[\frac{F_{q_1}}{\tn_0},\frac{G_{q_1}}{\tn_0}\right]-q_1[F_{D_1},G_{D_1}]\right)+O(\eps^0).
\eenn
Thus the dynamics  emerges to leading order at  $\eps^{-1}$ (which gives the dynamics on a time-scale of order $\eps$) and to  next order at  $\eps^0$ (whose influence happens on a time-scale of order one).

 First we study the dynamics given by the leading order. It should be noticed that $q_1$ is constant, since the leading order Poisson bracket does not contain any functional derivatives with respect to $q_1$, and that
\benn
&& \dot{\tn}_1=\frac{1}{\eps^2}\lap H_{D_1}+O(\eps)=-\frac{\tn_0}{\eps}D_1+[\tn_1,\ilap(q_1\tn_0+q_0\tn_1)]- \nabla\cdot (\tn_1\nabla\ilap D_1)+O(\eps),\\
&& \dot{D_1}=-\frac{1}{\eps^2}\lap H_{\tn_1}+\frac{1}{\eps}\nabla\cdot \left(\frac{H_{q_1}}{\tn_0}\nabla q_1\right)+\frac{1}{\eps}[H_{D_1},q_1]+O(\eps).
\eenn
If we impose the following constraints on the initial conditions:
\benn
&& \lap H_{\tn_1}=0,\\
&& \lap H_{D_1}=0,
\eenn
then these constraints are preserved by the leading order flow.
These constraints are equivalent to the following:
\benn
&& D_1=0,\\
&& \frac{T}{M}\frac{\lap \tn_1}{\tn_0}-q_0 \tn_0(q_0 \tn_1+q_1 \tn_0)=0.
\eenn
Note that, from expanding $n({\bf x},t)=n_0(\eps {\bf x})+\eps n_1({\bf x},t)$ about $\eps =0$, it follows that $\delta n_0$ is a linar function of ${\bf x}$ and, as a consequence, it does not appear in the equations for the constraints. A generalization to the case of non-harmonic $\delta n_0$ is however possible.

Even if we neglect  higher order terms ($\eps^2$ in the Hamiltonian and $\eps^0$ in the Poisson bracket), these constraints are not preserved by the Poisson bracket. Therefore, these quantities are approximately conserved on a time scale of order $\eps$.
Next, we approximate the dynamics on a time-scale of order 1 by inserting the constraints on $D_1$ and $n_1$ into the second order Poisson bracket. By dropping all   dependence on $D_1$ and $n_1$, the dynamics is thus equivalently given by the Hamiltonian
$$
H_1= \int d^2x \frac{\tn_0}{2}\left(-(q_0 \tn_1+q_1 \tn_0)\ilap (q_0 \tn_1+q_1 \tn_0)+\frac{T}{M}\frac{\tn_1^2-2\tn_1 \delta n_0}{\tn_0^2}\right),
$$
where $\tn_1$ is a function of $q_1$ given by
$$
\tn_1=-\frac{\tn_0}{q_0}\left(1-\frac{T}{M\omega_c^2}\lap \right)^{-1} q_1,
$$
and the Poisson bracket 
\beq
\label{eq:HMrpb}
\{F,G\}_1= \int d^2x \, q_1\left[\frac{F_{q_1}}{\tn_0},\frac{G_{q_1}}{\tn_0}\right],
\eeq
which satisfies the properties of a Poisson bracket - in particular,  the Jacobi identity.
Using this condition on $n_1$, the Hamiltonian $H_1$ can be rewritten as
\beq
\label{eq:HMrh}
H_1=\frac{T\tn_0}{2Mq_0^2}\int d^2x \left[q_1 \left(1-\frac{T}{M\omega_c^2}\lap \right)^{-1} q_1 -2\lambda \left(1-\frac{T}{M\omega_c^2}\lap \right)^{-1} q_1\right],
\eeq
where $\lambda$ contains the spatial variations of the equilibrium density as follows: 
$$
\lambda=-\frac{q_0}{n_0}\delta n_0({\bf x}).
$$
Using the symmetry of the operator $(1-(T/M\omega_c^2)\lap)^{-1}$, the Hamiltonian~(\ref{eq:HMrh}) can be rewritten as
\beq
\label{eq:HMrh2}
H_1=\frac{T\tn_0}{2Mq_0^2}\int d^2x (q_1-\lambda) \left(1-\frac{T}{M\omega_c^2}\lap \right)^{-1} (q_1-\lambda).
\eeq
Up to some constants, the Poisson bracket~(\ref{eq:HMrpb}) and the Hamiltonian~(\ref{eq:HMrh2}) are indeed the same as those presented  in Ref.~\cite{wein83}. Thus we have provided a derivation process that leads to  dynamics, on time-scales of order 1, that is still generated by a Hamiltonian and a Poisson bracket.

\section{Conclusion}

An important issue in the derivation of reduced models for plasma physics is avoiding the introduction of fake dissipative terms, which may result from uncontrolled approximations and truncations in the derivation process. In particular, if  the parent model has a Hamiltonian structure, we argue that the final reduced model should   also have a Hamiltonian structure.

 In this paper we examined, in this spirit, the case of the CHM equation. In particular we showed how the fundamental elements, i.e.,  the Hamiltonian functional and the Poisson bracket, of the Hamiltonian formulation of the CHM equation,  emerge from the Hamiltonian structure of a parent model, which is the starting point of the derivation commonly adopted in the plasma physics literature. The appearance of the Hamiltonian and the bracket of the CHM equation in the derivation process was seen to be facilitated  by adopting the new set of variables $(q,\tilde{n},D)$. In terms of these variables, the part of the bracket of the parent model that becomes  the CHM bracket can be easily identified. Indeed, what our paper shows is how the ordering adopted in the derivation  is able to reduce the bracket of the parent model to the CHM bracket, without compromising the fundamental properties of a Poisson bracket, such as for instance
  the Jacobi identity. A further new element of our analysis is the way the plasma compressibility is treated. Without invoking the drift approximation and the polarization drift, the divergence-free condition on the plasma velocity appears as a solution for the variable $D_1$ on a time scale of order $\epsilon$. Such a solution is used in order to approximate the dynamics of order $1$, assuming  for such dynamics that an  equilibrium solution exists. A similar argument is used for the dynamics of $\tilde{n}_1$, which  at the lowest order  is constrained to be a function of $q_1$ or, more precisely, to be proportional to the plasma stream function.

 We believe that the method adopted in this paper is a framework for deriving the Hamiltonian structure in other reduced models of plasma physics, and for deriving  new models while avoiding  the risk of introducing fake dissipative terms.

\acknowledgments
This work was supported by the European Community under the contract of Association
between EURATOM, CEA, and the French Research
Federation for fusion studies. The views and opinions expressed herein do not
necessarily reflect those of the European Commission. Financial support was also received from the
Agence Nationale de la Recherche (ANR EGYPT) and PJM was
supported by the US Department of Energy Contract No.~DE-FG03-96ER-54346. We acknowledge useful discussions with Alain Brizard and the Nonlinear Dynamics group at the Centre de Physique Th\'eorique.

\end{document}